# Fractal triangular search: a metaheuristic for image content search

*Erick O. Rodrigues[1,2] ✉, Panos Liatsis[3], Luiz Satoru[1], Aura Conci[1]*

*[1]Department of Computer Science, Universidade Federal Fluminense, Niteroi, Rio de Janeiro, Brazil*

*[2]Department of Computer Science, Universidade Federal de Itajuba (UNIFEI), Itabira - Minas Gerais, Brazil*

*[3]Deparment of Computer Science, Khalifa University of Science and Technology, Abu Dhabi, United Arab Emirates*

✉ *E-mail: erickr@id.uff.br*

**Abstract:** This work proposes a variable neighbourhood search (FTS) that uses a fractal-based local search primarily designed for images. Searching for specific content in images is posed as an optimisation problem, where evidence elements are expected to be present. Evidence elements improve the odds of finding the desired content and are closely associated to it in terms of spatial location. The proposed local search algorithm follows the fashion of a chain of triangles that engulf each other and grow indefinitely in a fractal fashion, while their orientation varies in each iteration. The authors carried out an extensive set of experiments, which confirmed that FTS outperforms state-of-the-art metaheuristics. On average, FTS was able to locate content faster, visiting less incorrect image locations. In the first group of experiments, FTS was faster in seven out of nine cases, being >8% faster on average, when compared to the second best search method. In the second group, FTS was faster in six out of seven cases, and it was >22% faster on average when compared to the approach ranked second best. FTS tends to outperform other metaheuristics substantially as the size of the image increases.

## 1 Introduction

Searching for patterns and content in images is a common problem in various applications of visual computing, such as image registration [1, 2], content tracking and pattern recognition [3, 4], 3D reconstruction and geometry estimation [5], image data and feature analysis [4, 6, 7]. Nevertheless, the search problem is often addressed from very different perspectives.

In image registration, a template is displaced over an image, where at each possible position, a similarity score is computed [7]. Template matching, for instance, is fundamentally a search problem, where similar regions in relation to the template, according to a similarity measure, are searched for. However, in most cases, no intelligent heuristic is used to improve the performance of these algorithms. In other words, the most commonly used paradigm is exhaustive search, where every position or candidate region in the image is considered.

This work contributes to the fields of visual computing and optimisation by (i) defining a search problem in images called search based on evidence (SBE), (ii) defining rules for generating instances for further comparison, and (iii) proposing a robust variable neighbourhood search (VNS) [8] metaheuristic based on a fractal local search strategy [9] called fractal triangular search (FTS) to solve the SBE problem. Extensive experiments were performed to analyse the efficiency of the proposed algorithm, in comparison to state-of-the-art methods.

Searching for features of the human face is a practical example of SBE. Let us suppose that the location of the human nose is to be determined in a sequence of images of human faces. The nose could be located anywhere in the image. Exhaustive search requires consideration of every possible image position, and finally, the most similar region is selected. SBE would model this process using evidence data. For instance, the eyes are always close to the nose, and therefore, this information can be used to speed up the search. If the eyes are detected prior to the nose, then a priori content related to the nose will be present nearby, and this fact can bias the behaviour of the search, triggering, for instance, a specific local search. Approaches such as the example described can locate a specific piece of content, e.g. nose, faster on average. Image regions that correspond to the nose or the eyes can be considered using templates of the expected regions coupled with similarity measures.

The metaheuristic proposed in this work can be applied in images or 2D discrete spaces. Motion estimation is a potential application, where the metaheuristic can accelerate the block matching process. A hypothetical application is searching for an object that is displaced in the frames of a video sequence. By supposing that the object needs to be searched throughout the spatial extent of a frame, since its location cannot be predicted in advance, evidence elements associated to the object could be used to speed up the search. Locating an evidence element indicates that the main blocks are spatially close, and local search can be forced, potentially inducing faster convergence. This concept can be applied in content-based image search/retrieval. Image content must be searched for, and hence FTS can also be used to speed up the search.

This paper is organised as follows: In the next section, a review of the state of the art and some details of the proposed approach are presented. In Section 3, SBE is formally defined. Next, various algorithms to solve the SBE problem are presented. Section 4 provides the comparison and analyses the obtained results. Finally, the final section describes the conclusions of this research and provides directions for future work.

## 2 Literature review

Metaheuristics have been commonly used for parameter optimisation in visual computing methods. Jiang *et al.* [10] proposed an evolutionary tabu search (TS) for optimising the parameters of an ellipse, with the goal of autonomously segmenting cells. In a previous work, we employed a similar idea to segment the pericardium layer of the human heart [11]. Parameter optimisation has also been used to reduce the computational complexity in image registration [12, 13]. Furthermore, it can also be used in feature selection [14], pattern recognition [15, 16] etc.

Two images are usually involved in image registration. The goal is to find an optimal or an appropriate near optimal transformation of one image causing it to resemble the whole or part of the other [1]. In regards to pattern recognition, the goal is to align landmarks between two images [17] or to use similarity measures to find the most suitable position of a template [18, 19]. 3D reconstruction and geometry estimation can also involve the use of search for specific areas and patterns [20]. In medical applications, for instance, several approaches extract features from specific areas of images (that could be searched for) for providing support data or diagnosis [4, 7].

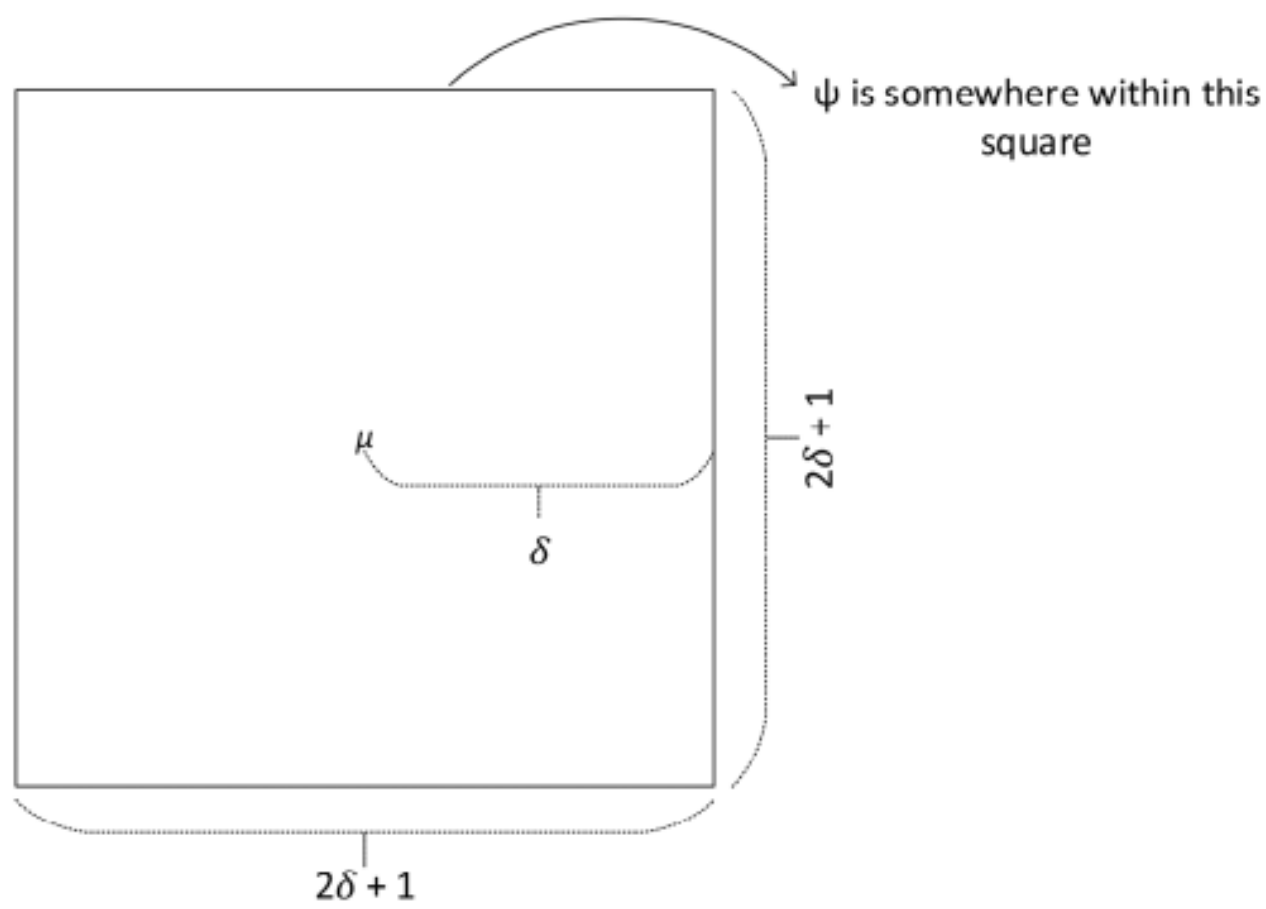

**Fig. 1** *Relationship between μ and ψ*

Besides belonging to the field of visual computing, all these problems have a common characteristic. That is, the aforementioned applications can be posed as optimisation problems that consist of searching for a visual pattern (i.e. template) within images, which is usually supported by surrounding evidence, suggesting that the searched element is nearby (e.g. a certain pattern, pixel, area of pixels etc.). We define the evidence element as any data within the image that can be useful in improving the efficiency or the odds of finding the desired content.

The majority of research works in visual computing do not make use of evidence information in the context of image content search. This work proposes the SBE paradigm, which uses evidence information to bias the search for image content, and proposes a metaheuristic that can efficiently locate the global optima (or the near global optima, if desired) for SBE problems. In this context, the global optima is equivalent to the exact content or searched region.

In pattern recognition and registration, SBE can be used to speed up the matching of points. In the literature, some image segmentation approaches are based on machine learning methods [21]. Among these, some are based on classification algorithms [7, 22]. In some segmentation methods, only selected parts of an image are segmented, and therefore, these parts can be searched for using SBE, thus improving performance. Several works, mainly in the field of visual computing, could directly benefit from the use of SBE problem solvers.

In order to validate the proposed metaheuristic, we compare FTS to three variations of the VNS, iterated local search (ILS) [23], TS [24] and exhaustive search. Evolutionary algorithms [25] were used to optimise the parameters of each metaheuristic (FTS, VNS, ILS and TS). Exhaustive search is not a metaheuristic itself but rather a brute force method, where each image location is considered. A large number of experiments were performed, and the average-case step complexity was analysed and compared.

## 3 Materials and methods

Let $\boldsymbol{M}$ be a matrix, where $M_{i,j}$ represents the element at row $i$ and column $j$, such that $j, i \in \mathbb{Z}^2$. The searched element or content is defined as $\psi$ (as previously explained, this could be a region, a linear structure, a pixel intensity etc.). Here, $\psi$ is an element of the matrix, such that $\psi = M_{\hat{i},\hat{j}}$, for a single possible combination of $\hat{i}$ and $\hat{j}$. Evidence elements are defined as $\mu$ such that the set $\{(\tilde{j},\tilde{i}) : M_{\tilde{i},\tilde{j}} = \mu\}$ contains the coordinates of all evidence elements, and its cardinality must be greater than 0. $(\tilde{j},\tilde{i})$ assumes all possible coordinates.

Formally, $\rho(\psi \mid \mu)$ must be greater than $\rho(\psi)$. That is, the probability of finding $\psi$ given any evidence $\mu$ must be greater than the probability of just finding $\psi$ with no evidence. As $\psi$ is nearby evidence elements $\mu$, this information can be used to search for $\psi$ in a limited area around $\mu$. If $\mu$ is always within a maximum distance $\delta$ from $\psi$, then the worst-case scenario for finding $\psi$ is $(\delta + 1)^2$ steps, which corresponds to the square area around $\mu$.

Definition 1 formulates the SBE problem and Theorem 1 proves that the chance of finding $\psi$ given at least one element $\mu$ in an SBE problem is higher than without any knowledge of $\mu$.

*Definition 1:* SBE problem
Variable $\psi$ represents the searched element/content in matrix $\boldsymbol{M}$, where $\psi = M_{\hat{i},\hat{j}}$ for a single combination of $\hat{i}$ and $\hat{j}$. $\Omega$ stands for a set that contains the locations $(\tilde{j},\tilde{i})$ of evidence elements $\mu$. $P$ represents a set that contains the locations $(i, j)$ of matrix $\boldsymbol{M}$. $\delta$ corresponds to the maximum distance between all elements $\mu$ and $\psi$, and $\boldsymbol{w}$ and $\boldsymbol{h}$ stand for the width and height of the matrix, respectively. $d(p,q)$ computes the Chebyshev distance between points $p$ and $q$. Finally, $\rho(\psi)$ represents the position $(\hat{i},\hat{j})$ of the searched content $\rho$. Given these definitions, the five following conditions must be satisfied:

$$① \ \mathrm{d}(\psi_{\mathrm{p}}, q) \le \delta \quad \forall q \in \Omega, \psi_{\mathrm{p}} = (\hat{i},\hat{j})$$

$$② \ \Omega \subset P$$

$$③ |\Omega| > 0$$

$$④ \ \min(\boldsymbol{w},\boldsymbol{h}) > 2\delta + 1 (\text{see Theorem 1 for reasoning})$$

*Theorem 1:* $\rho(\psi \mid \mu) > \rho(\psi)$ for any SBE problem. In other words, the conditional probability of randomly locating $\psi$, given that at least a single $\mu$ is known, is greater than the marginal probability of randomly locating $\psi$ for any problem that adheres to Definition 1.
The probability of randomly locating $\psi$ in $\boldsymbol{M}$ is

$$\rho(\psi) = \frac{1}{\boldsymbol{wh}}$$

where $\boldsymbol{w}$ and $\boldsymbol{h}$ represent the width and height of matrix $\boldsymbol{M}$, respectively.
Given an arbitrary $\mu$, it is known by Definition 1.① that $\psi$ is located at distance $\delta$ or less in relation to $\mu$. The area around $\mu$ is given by $(2\delta + 1) \times (2\delta + 1)$, as shown in (Fig. 1).
The probability of finding $\psi$ in this limited extent area is defined as

$$\rho(\psi \mid \mu) = \frac{1}{(2\delta + 1) \times (2\delta + 1)} = \frac{1}{4\delta^2 + 4\delta + 1}$$

Let us assume a square matrix$\boldsymbol{w} = \boldsymbol{h} = \boldsymbol{s}$. For $\rho(\psi \mid \mu) > \rho(\psi)$ to be true, the following inequality must hold:

$$\frac{1}{4\delta^2 + 4\delta + 1} > \frac{1}{s^2}$$

As $s$ is always >0, then:

$$\frac{1}{4\delta^2 + 4\delta + 1} > \frac{1}{s^2}$$

$$\vdots$$

$$s > 2\delta + 1$$

Thus, it is clear that as long as $s > 2\delta + 1$, or $\min(\boldsymbol{w},\boldsymbol{h}) > 2\delta + 1$, the inequality $\rho(\psi \mid \mu) > \rho(\psi)$ holds.

As a remark, the lower the $\delta$ variable in Theorem 1, the greater is $\rho(\psi \mid \mu)$ in relation to $\rho(\psi)$.

1 **method** $P$ generateInstance($s$);
2 **begin**
3 $M \leftarrow$ Generate a matrix or image of $s \times s$ size;
4 Place $\psi$ at a random position in $M$, $M_{i,j} = \psi$, where $(i, j)$ is randomly generated;
5 Place $n$ elements $\mu$ in $M \mid L_\infty(\psi_p, \mu_p) \leq s/10 \; \forall \mu$, $n = s/16$;
6 return $M$;
7 **end**

**Fig. 2** *Algorithm 1: instance generation algorithm*

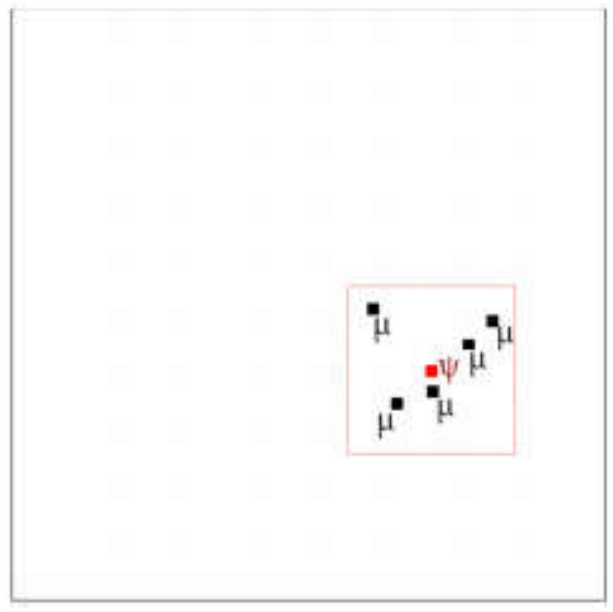

**Fig. 3** *Possible instance outcome*

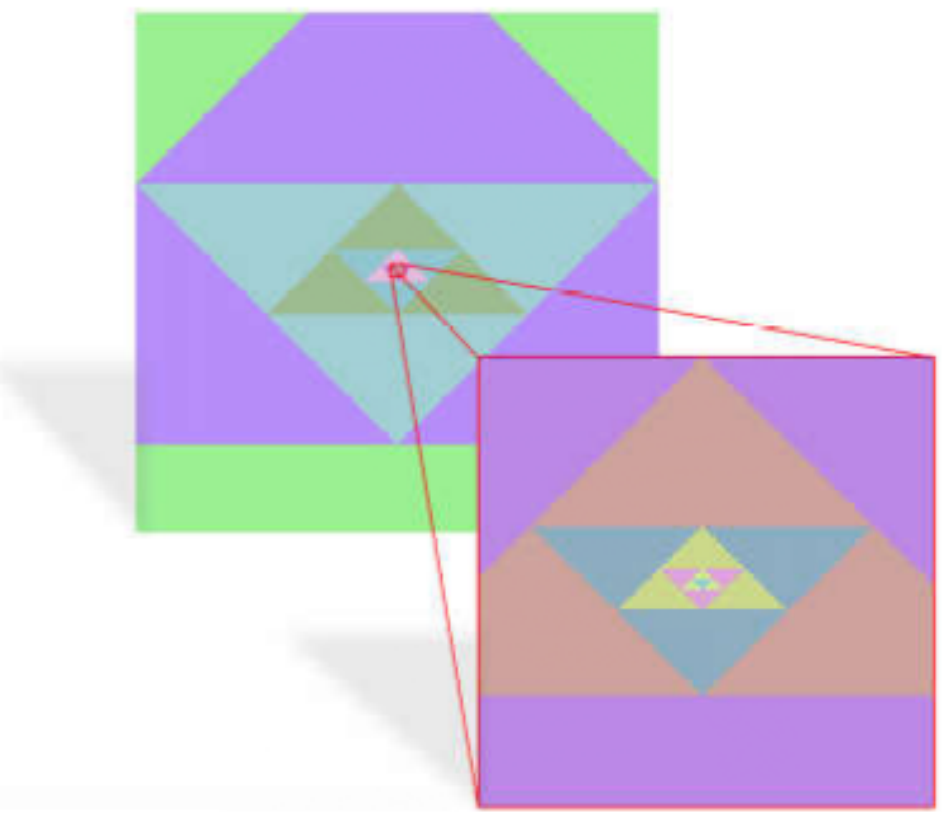

**Fig. 4** *Overview of the fractal triangular growth process*

### *3.1 Instance generation*

For the average-case experiments, synthetic instances were created in real time. In this subsection, we describe this process. The function that generates real-time instances receives $s$, which is equal to the width and height of the image. Thereafter, the $\psi$ value is randomly placed at an image location and a total of $s/16$ pixels, representing $\mu$ are randomly placed, respecting the SBE restrictions, at a maximum distance of $s/10$ pixels away from $\psi$, when considering the Chebyshev distance metric.

*Proof:* Created instances conform to the SBE problem.
We need to prove that $\min(\boldsymbol{w}, \boldsymbol{h}) > 2\delta + 1$ holds for this case. As $\boldsymbol{w} = \boldsymbol{h} = \boldsymbol{s}$ and $\delta = \boldsymbol{s}/10$

$$\boldsymbol{s} > 2\boldsymbol{s}/10 + 1$$
$$\vdots$$
$$\boldsymbol{s} > \boldsymbol{s}/5 + 1$$

So, if $\boldsymbol{s}$ is greater than $\boldsymbol{s}/5 + 1$, the last SBE condition is satisfied. The smallest $\boldsymbol{s}$ for this work was $\boldsymbol{s} = 1024$. Therefore, condition ④ is clearly satisfied. The remaining conditions are trivially satisfied, given the instance description. □

Algorithm 1 (see Fig. 2) illustrates the steps to generate a single instance, where function $L_\infty$ computes the Chebyshev distance, shown in (1), and $\boldsymbol{s}$ is again the height and width of the matrix, assuming points $q_1 = (x_1, y_1)$ and $q_2 = (x_2, y_2)$

$$L_\infty(q_1, q_2) = \max(|x_2 - x_1|, |y_2 - y_1|) \tag{1}$$

Fig. 3 shows a potential instance generated by Algorithm 1 (Fig. 2), where the black square outline represents the entire image $\boldsymbol{M}$, and the dashed red square represents the maximum $\delta$ distance from $\psi$, where the elements $\mu$ could be placed.

In summary, all algorithms in this work have the ultimate goal of finding $\psi$. Therefore, at each pixel or element evaluation, they repeat the process of checking whether the element is $\psi$. Once found, the algorithm converges. Otherwise, the algorithm verifies if it is $\mu$. If it is equal to $\mu$, the algorithm biases the search, as there is evidence that $\psi$ is spatially close. We define function visit($e$) as shown in (2) to represent this process.

Function visit($e$) examines element $e$ and provides feedback to the algorithm to converge if the searched element $\psi$ is found ($e = \psi$). Otherwise, it returns true if an evidence element $\mu$ is found ($e = \mu$). If this is not the case, it returns false. Function visit is used throughout this work in all search algorithms with the exception of the exhaustive search. The exhaustive search algorithm searches for $\psi$ without considering the presence of evidence $\mu$, and therefore this function is not required

$$\text{visit}(e) = \begin{cases} \text{converges} & \text{if } e = \psi \\ \text{returns true} & \text{if } e = \mu \\ \text{returns false} & \text{otherwise} \end{cases} \tag{2}$$

The essence of metaheuristics is that they do not always provide the best solution, i.e. the global optima. However, in the experiments reported in this work, all of the tested techniques were run until the specific element was visited (global optima). Thus, we count the number of elements visited before finding element $\psi$. This number is defined as the average-case time or step complexity, in practical situations.

### *3.2 Overview of the fractal local search*

The FTS metaheuristic is based on fractals [9]. More specifically, the method is a variation of the VNS, where the local neighbourhood search relies on a fractal growth concept. The premise is to generate several small triangles within the image and to increase their size in a fractal fashion until a convergence threshold is reached. The threshold condition is satisfied when evidence $\mu$, or the searched element $\psi$ are found. The fractal neighbourhood concept is shown in Fig. 4, where different colours illustrate the elements visited at each iteration.

Triangular patterns were chosen as they can make up any other existing polygon if arranged in an appropriate fashion, including concave polygons. Furthermore, a remarkable advantage that triangles contribute to local search methods is due to their irregular shape. That is, the area of the base is larger than the area of the top of the triangle, which can be exploited in search algorithms.

Fig. 5 shows an arbitrary iteration of the proposed algorithm. Numbers 1, 2 and 3 indicate the order in which elements are iterated in the local search. At first, the algorithm starts by iterating triangle 1 and its elements. Assuming that the desired content is not found during the iteration, triangles 2 and 3 are subsequently iterated. The figure illustrates how triangles can be appropriately exploited in a search problem. Triangles 2 and 3 make up the base of the outer large triangle formed by triangles 1, 2, 3 and the central area that corresponds to the previous iteration. The area of triangles 2 and 3, together, includes more pixels than the area encapsulated by triangle 1. Therefore, it is possible to argue that, once all these triangles have been iterated, the element or evidence searched for is most probably closer to triangle 1 rather than to triangles 2 and 3, as a larger number of pixels near the bottom of the image have been already checked.

After iterating triangles 2 and 3, the algorithm inverts the triangular logic over the horizontal axis, as shown in Fig. 6. As it can be noted, triangles 5, 6 and 7 are the triangles that will be subsequently iterated. In the following iteration, a wider area on the top of the image is checked to the detriment of a wider area at the bottom, as the previous iteration indicated that elements may be closer to the top rather than the bottom.

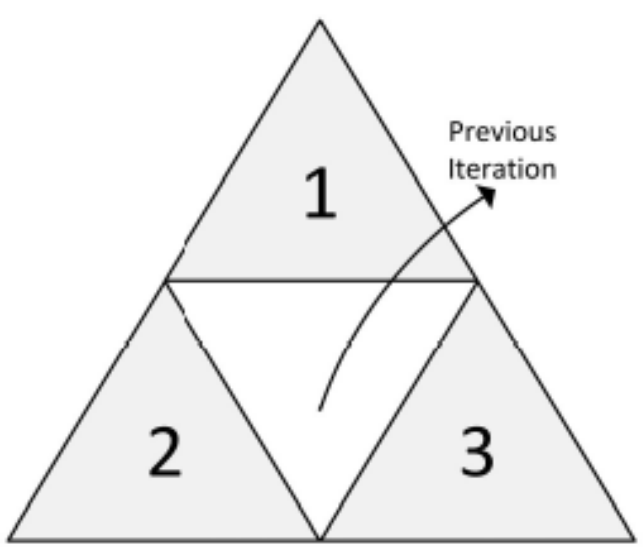

**Fig. 5** *Triangle disposition at iteration n*

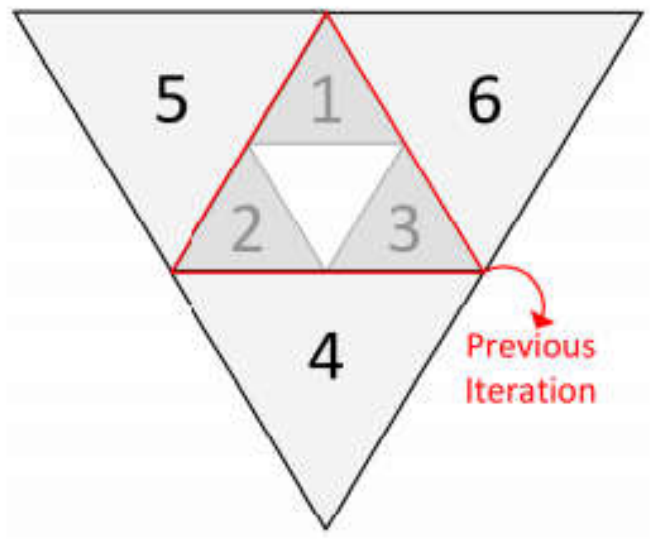

**Fig. 6** *Triangle disposition at iteration $n + 1$*

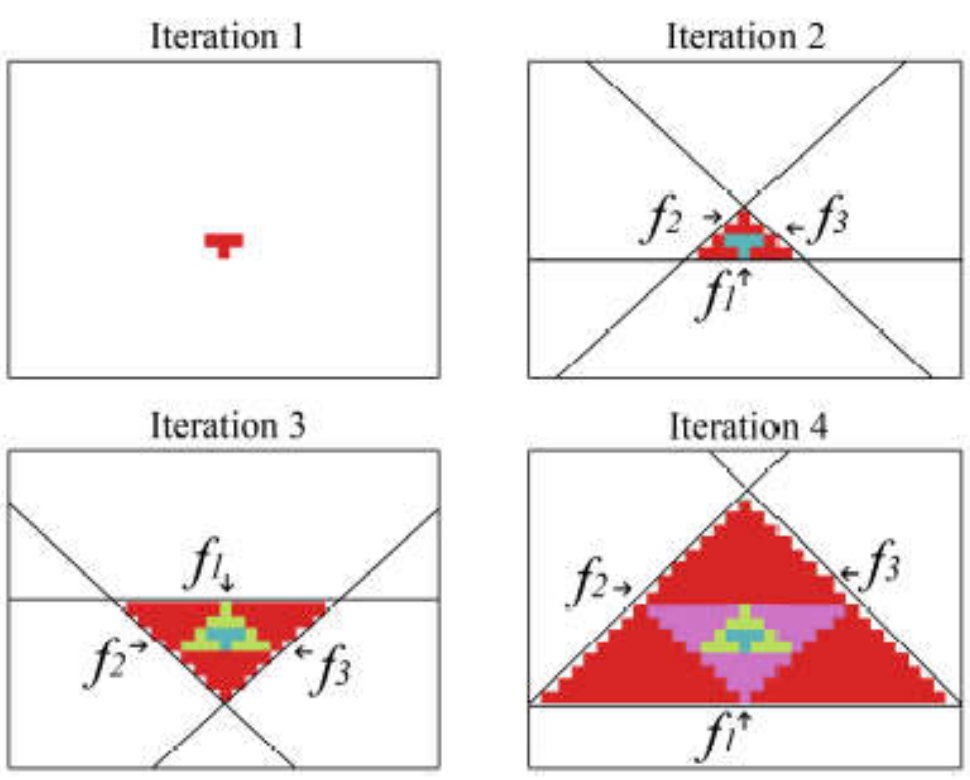

**Fig. 7** *Steps of the neighbourhood evaluation*

The logic described in Fig. 4 and Fig. 6 is respected throughout the iterations of the algorithm. This methodology differs from other local searches, avoiding the use of symmetric patterns in iterated spaces. Furthermore, we propose a very efficient algorithm that generates these triangles and iterates each of their elements without revisiting elements, avoiding the use of a tabu list.

### 3.3 Fractal triangular search

In what follows, we shift our attention to a finer explanation of the proposed methodology and related algorithms. At first, a small discrete triangle encapsulating four elements (or pixels) is placed somewhere in the image. Let $f_1, f_2, f_3$ be functions that tightly segregate all the elements in $S$. Fig. 7 illustrates this process. The red area represents elements that are visited and evaluated at each iteration.

The approach consists of generating several of the triangles shown in Fig. 7. Next, they should be grown at each iteration while their pixels, or elements within, are visited and evaluated. A total of $t$ triangles are generated within the image at distance $d$ from each other. Given the $(x, y)$ central position of a generated fractal triangle, the subsequently generated triangle must be at the position given by (3), where $x$ and $y$ are the coordinates of the former triangle, $\boldsymbol{w}$ and $\boldsymbol{h}$ represent the width and height of the image, respectively, $d$ is the distance between both triangles, on the $x$ and $y$ axes, respectively, % represents the modulo operation (remainder of the division), and $\gamma$ randomly assumes true or false with equal probability at each access of this variable. The triangles are generated and processed in a queue fashion. The distance requirement has to be true for every consecutive pair of triangles in the queue

$$\mathrm{pos}(x, y) = ((x + \Delta_x)\%\boldsymbol{w}, (y + \Delta_y)\%\boldsymbol{h})|$$
$$\Delta_x = \begin{cases} d & \text{if } \gamma = \text{true} \\ 0 & \text{otherwise} \end{cases}, \quad \Delta_y = \begin{cases} d & \text{if } \gamma = \text{false} \\ 0 & \text{otherwise} \end{cases} \tag{3}$$

Each generated triangle iterates and grows its neighbourhood $c$ times. If any evidence is found, the fractal growth pattern continues until the searched element $\psi$ is reached. If neither evidence nor the searched element is found and the generated triangles have been grown for $c$ times, then the algorithm performs an exhaustive search. The worst-case scenario of all the metaheuristics in this work is the exhaustive search, which has a complexity of $O(s^2)$. However, the purpose of the current investigation is to determine how the algorithms perform on average, rather than evaluate the worst-case scenario.

Parameters $t$, $d$ and $c$ compose the input data, as shown in the flowchart of Fig. 8, which demonstrates the steps and rationale of the algorithm.

Fig. 9 shows a possible scenario where the little rhombus represents the evidence elements and the circle represents the searched element. For this specific case, the set of input parameters has to be $t \geq 3$, $d$ and $c = 2$. The first triangle is generated at the top-left position of the image matrix. The algorithm iterates this triangle twice ($c = 2$, the two colours of the triangle), since no evidence is found. A second triangle is generated at a distance $d$ from the first one. The same process is repeated. A third triangle is generated at distance $d$ from the second triangle. In this third triangle, an evidence element is found at the second iteration and therefore the triangle keeps growing until it finds the searched element $\psi$, which, in this case, is at the third iteration (three colours – the third iteration shown in green).

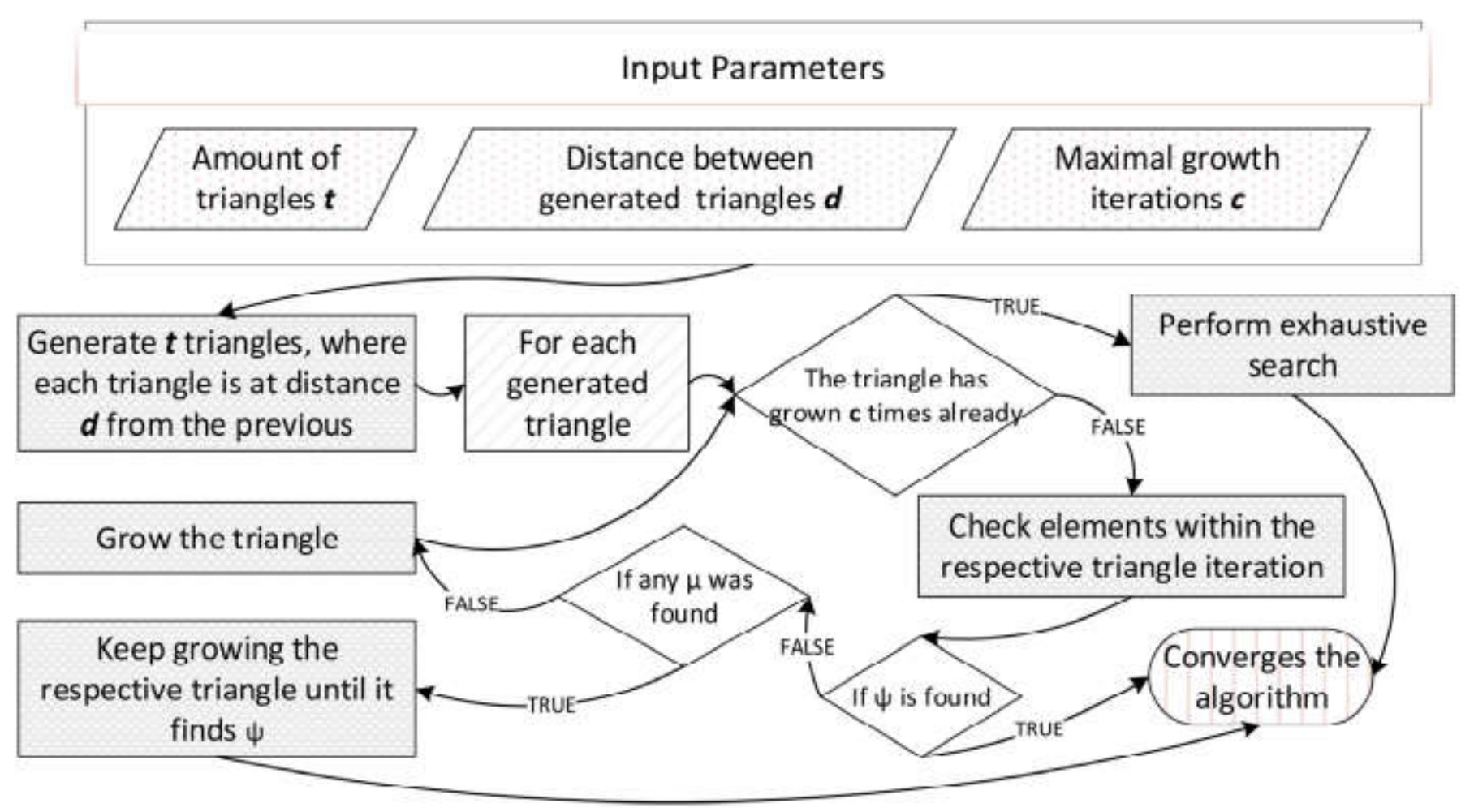

**Fig. 8** *Steps of the FTS algorithm*

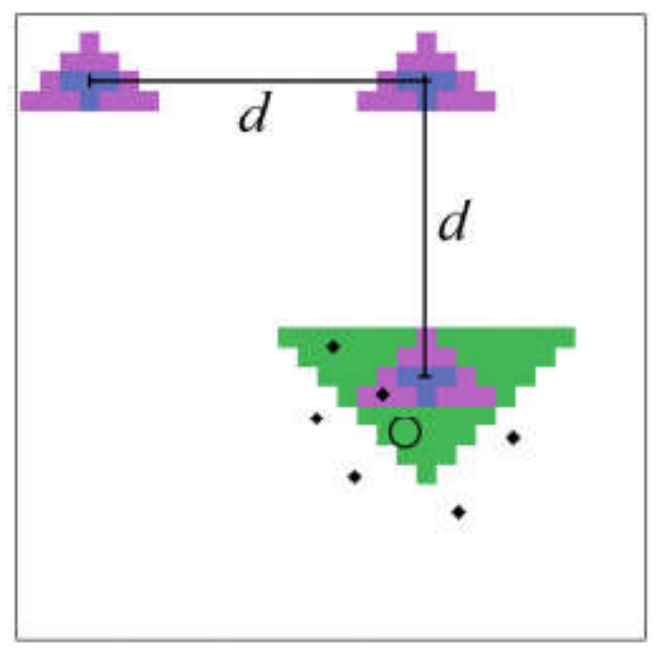

**Fig. 9** *Overall view of the FTS metaheuristic execution*

```
1  method boolean triangleStretch(x, y, y_c, h);
2  begin
3  |  if (visit(M_{y,x})) return true;
4  |  x_aux ← x;
5  |  for k_h = 1...h do
6  |  |  y ← y + y_c;
7  |  |  x ← x − 1;
8  |  |  x_aux ← x_aux + 1;
9  |  |  for l = x...x_aux do
10 |  |  |  if (visit(M_{y,l})) return true;
11 |  |  end
12 |  end
13 |  return false;
14 end
```

**Fig. 10** *Algorithm 2: triangle iteration*

Clearly, it is more efficient to not revisit elements that have been already visited, preferably without using a tabu list [26]. Producing this effect in a discrete domain using the functions $f_1$, $f_2$ and $f_3$ of Fig. 7 would be computationally expensive. Thus, we developed a simple and robust discrete solution shown in Algorithm 3 (Fig. 11). Algorithm 2 (Fig. 10), which concerns to the triangle iteration, is part of Algorithm 3 (Fig. 11). Algorithm 2 (Fig. 10) iterates triangles given a central position $(x, y)$, an orientation $y_c$ and height $h$.

The effect produced with Algorithm 2 (Fig. 10) is not equivalent to a tabu list and therefore it is not meant to replace it. For instance, two triangles may end up visiting the same area, and both will eventually revisit previously visited elements. However, Algorithm 2 ensures that at least for a single triangle, the elements are not revisited (Figs. 10 and 11).

### 3.4 Iterated local search

The ILS approach divides the image in $t$ squares of equal size. A certain number $a$ of elements is visited within each square. If an evidence element is found, all the elements in the square are visited. The perturbation phase occurs when the algorithm switches from one square to another. The ILS pseudo-code is shown in Algorithm 4 (Fig. 12).

### 3.5 Variable neighbourhood searches

As previously mentioned, FTS is a type of VNS algorithm. Besides FTS, three other versions of VNS are analysed, as they are the most similar in relation to FTS. The first VNS approach is called VNS1. VNS1 consists of selecting $t$ random elements at a distance $d$ from each other. If an evidence $\mu$ is found, the algorithm iterates through a neighbourhood of size $m$ around the found evidence element. The first position is always generated completely at random. The following coordinates are given by 3. The pseudo-code in Algorithm 5 (Fig. 13) illustrates the core idea of VNS1. It is important to highlight that this neighbourhood structure or local search is much simpler than the one in the FTS algorithm.

The second VNS algorithm (VNS2) is slightly different. The neighbourhood size is not fixed. Parameter $m$ is not present in VNS2. If an evidence element is found, the neighbourhood of that element is visited in a growing fashion. That is, the surrounding elements are visited, respecting the $L_\infty = \max(|x_j - x_c|, |y_i - y_c|)$ order, from the closest to the farthest, where $(x_j, y_i)$ represent the coordinates of the iterated neighbouring element and $(x_c, y_c)$ is the coordinate of the central evidence element $\mu$.

At last, VNS3 is the method with the greater similarity to the proposed FTS algorithm. The difference lies in the neighbourhood structure/local search. In this case, in the first iteration, a square of few elements is generated. The square is visited and grown $g$ times, where its neighbourhood is visited just like in VNS2. If no evidence is found during $g$ iterations, the algorithm generates another square at $d$ distance from the former and repeats the process. If any evidence is found then the neighbourhood is grown until the searched element is found. The total amount of generated squares is controlled by $t$. In the worst case, where no evidence and search elements are found, an exhaustive search is performed. VNS3 is described in Algorithm 6 (Fig. 14).

### 3.6 Tabu search

TS is essentially the previously described VNS2 approach with the use of a tabu list. That is, at the beginning, the algorithm initialises a Boolean matrix of the size of the input matrix. Once each element is visited, it is marked as already visited in the Boolean matrix. If it is eventually visited again, this does not count as an access.

The reason for using the tabu list with VNS2 instead of VNS3 is so that the similarities of VNS3 in relation to FTS are preserved. Therefore, using a tabu list with VNS3 would be unfair as FTS does not use a tabu list either. A fair comparison would be either FTS + tabu and VNS3 + tabu or none of them using the tabu list. Tabu lists add a further computational burden, due to the processing of potentially massive search spaces. Therefore, it was

```
Data: s are equal to the width and height of the input matrix
      M, t is the total number of generated triangles, d is the
      distance between the triangles, c is the convergence
      parameter and random(b1, b2) is a function that generates
      a random integer between b1 and b2 (both inclusive).
1  method fractalTriangularSearch(t, d, c);
2  begin
3      x ←random(0, s − 1); y ←random(0, s − 1);
4      count ← 0; yc ← −1; h ← 2; w ← 3;
5      v ← triangleStretch(x, y, yc, h);
6      for each kt triangle do
7          while count < c | | v do
8              yc ← −1 ∗ yc;
9              count ← count + 1;
10             ceil ← ⌈w/2⌉;
11             v ← triangleStretch(x − ceil,
                 y + (−yc ∗ (h − 1)), yc, h);
12             v ← triangleStretch(x + ceil,
                 y + (−yc ∗ (h − 1)), yc, h);
13             v ← triangleStretch(x, y + (−yc ∗ (h ∗ 2 − 1)),
                 yc, h);
14             h ← 2 ∗ h;
15             w ← w ∗ 2 + 1;
16             y ← y − yc ∗ (h − 1);
17         end
18         (x, y) ← pos(x, y, d);
19         count ← 0; yc ← −1; h ← 2; w ← 3;
20         v ← triangleStretch(x, y, yc, h);
21     end
22     If ψ has not been found then perform
         exhaustiveSearch(M);
23 end
```

**Fig. 11** *Algorithm 3: FTS algorithm*

```
Data: t is the amount of squares to divide the input matrix M
      in, and a is the amount of elements to be evaluated
      within each square
1  method iteratedLocalSearch(t, a);
2  begin
3      Divide M in t squares of the same size;
4      for each square k in random order do
5          for a total of 'a' randomly chosen positions p do
6              if visit(Mp) then
7                  for each position pe of elements e ∈ k do
8                      visit(Mpe);
9                  end
10             end
11         end
12     end
13     If ψ has not been found then perform
         exhaustiveSearch(M);
14 end
```

**Fig. 12** *Algorithm 4: ILS algorithm*

decided that tabu lists will not be used in VNS3 and FTS, so as not to affect their real-time characteristic.

### *3.7 Parameters selection*

Choosing the set of input parameters for the algorithms based on trial and error lacks in terms of a systematic approach. Furthermore, there is a significant number of parameters to be evaluated, and varying the size of the instances alters the optimal set of parameters. Therefore, it was decided to use a metaheuristic to search for a good set of parameters for each case.

The parameters were selected using an evolutionary algorithm (EA). The overall steps of EA are shown in Fig. 15. At first, the algorithm randomly generates five individuals and places them in the initial population. Next, the best fit individual is selected and there is (i) a 5% chance of the algorithm doing nothing and essentially 'cloning' the individual (elitist strategy); (ii) 65% chance of performing cross-over with the $(\alpha\kappa)$th best fit individual, where $\alpha$ is the product of two random floats $\in [0, 1]$ and $\kappa$ is the maximum size of the population (which was set to 50); as well as (iii) a 30% chance of occurring mutations, where a random digit is changed to a number $\in [0, 9]$ with equal probability.

The fitness function is defined as the mean amount of visited elements (including the evidence ones) until the algorithm finds the searched element. That is, at each generated instance image and for each parameter selection, each algorithm is run 40 times and the fitness function is the average of these 40 attempts. This mitigates the bias of the stochastic behaviour. As the objective function consists of computing the number of visited elements before finding the searched element $\psi$, the objective function should be minimised.

The EA stopping condition occurs when (1) all individuals in the population are equal, or (2) 100 iterations with no improvement is reached, the best fit individual did not change, and there is at least two copies of the best individual in the population. It is important to highlight that each algorithm used the same version of the evolutionary algorithm for parameter selection.

## 4 Experiments and results

The experiments are based on the law of large numbers (LNN) [27], which is a theorem that guarantees the stabilisation of the mean of random outcomes (which are essentially limited in an interval) over a large amount of time or iterations. If a normal dice is thrown *ad infinitum*, the average of the outcomes would be equal to 3.5. In theory, assuming an image of size $n \times n$, the average evaluated elements in an exhaustive search (disregarding the evidence element $\mu$) over a sufficiently large amount of iterations tends to $n^2/2$. That is, we have $n \times n$ elements in the image, and $\psi$ can be placed anywhere ($\psi$ changes location at each instance generation), the average of visited elements over a sufficiently large period of time would be close to half of the amount of elements in the image, which is $n^2/2$.

In our formulation, each algorithm starts off with randomly chosen parameters stipulated by the EA. Once the EA converges, the results obtained up to this point are ignored. These are 'bad results', as the parameters were still adapting. If the EA does not converge after a certain number of iterations, all generated results are averaged, including the so-called bad results. In the case of the FTS algorithm, the corresponding EA had the least number of convergences. In other words, the results obtained by the FTS could indeed be better than the results reported in the current work. However, even limited by the performance of the EA, the FTS algorithm outperformed the remaining techniques.

In the first instance, 1024 × 1024 images were used. Each one of the metaheuristics was run for $\sim 2 \times 10^7$ times, where at each run a new instance of the problem was generated (see Algorithm 1, Fig. 2). Therefore, a total of $2 \times 10^7$ instances were evaluated by each algorithm.

Moreover, a multi-start approach was employed. The EA was restarted a total of 9 times to avoid entrapment in local minima during the parameter selection process. That is, one could argue that the selected parameters were the result of entrapment in local minima. In other words, the EA may have not selected a fairly good set of parameters.

The proposed EA algorithm is well constructed and was extensively tested. Besides, it was set to run for a sufficient amount of time with a fair mutation rate. However, just to safeguard the quality of the solution, as previously mentioned, the EA was restarted nine times. Thus, for each metaheuristic, there was a total of $9 \times (2 \times 10^7)$ runs. The obtained results are shown in Table 1, where each line represents a set of parameters determined by the EA for each of the algorithms, and the numbers in the table cells are the averaged results (the mean of $2 \times 10^7$ runs in each table cell).

Each table cell is the average of $2 \times 10^7$ instances. For example, the number 523,816 of the FTS column (first line) is the average amount of visited pixels or elements over $2 \times 10^7$ different input instances. These values can be interpreted as the practical average step complexity. The *Mean* line in the table indicates the mean of the 9 values in the same column, while the *Lowest* line indicates

**Data:** $t$ is the amount of randomly visited elements, $s$ is the width and height of the matrix, $m$ is the size of the evaluated neighbourhood, $d$ is the distance between the elements, and random($b_1, b_2$) is a function that generates a random integer between $b_1$ and $b_2$ (both inclusive).

1 **method** variableNeighbourhoodSearch1($t$, $m$, $d$);
2 **begin**
3 $(x, y) \leftarrow$ (random$(0, s-1)$,random$(0, s-1)$);
4 $c \leftarrow 0$;
5 **while** $c < t$ **do**
6 **if** *visit(*$M_{y,x}$*)* **then**
7 **for** *each p position at maximum distance* $m$ *from* $(x, y)$*, starting the p iterations at the top-left position and stopping at the bottom-right one* **do**
8 $visit(M_p)$;
9 **end**
10 **end**
11 $(x, y) \leftarrow pos(x, y, d)$;
12 $c \leftarrow c + 1$;
13 **end**
14 If $\psi$ has not been found then perform exhaustiveSearch($M$);
15 **end**

**Fig. 13** *Algorithm 5: VNS1 algorithm*

**Data:** $s$ is equal to the width and height of $M$, $t$ is the number of generated squares, $d$ is the distance between each pair of squares, $g$ is the convergence parameter and random($b_1, b_2$) is a function that generates a random integer between $b_1$ and $b_2$ (both inclusive).

1 **method** variableNeighbourhoodSearch3($t$, $d$, $g$);
2 **begin**
3 count $\leftarrow 1$; $x \leftarrow$ random$(0, s-1)$; $y \leftarrow$ random$(0, s-1)$;
4 $k_1 \leftarrow \{P_{(y-1,x-1)}, P_{(y-1,x)}, P_{(y-1,x+1)}, P_{(y,x-1)}, P_{(y,x)}, P_{(y,x+1)}, P_{(y+1,x-1)}, P_{(y+1,x)}, P_{(y+1,x+1)}\}$;
5 $v \leftarrow v \mid\mid visit(e), \forall e \in k_1$;
6 **for** *each square* $k_s$ *from* $k_1$ *to* $k_t$ **do**
7 **while** $count < g \mid\mid v$ **do**
8 $count \leftarrow count + 1$;
9 **for** $i \leftarrow -count$ *to* $i \leftarrow count$ **do**
10 **for** $j \leftarrow -count$ *to* $j \leftarrow count$ **do**
11 **if** $(|i| = count) \mid\mid (|j| = count)$ **then**
12 $v \leftarrow visit(P_{i+y,j+x})$;
13 **end**
14 **end**
15 **end**
16 **end**
17 count $\leftarrow 1$; $(x, y) \leftarrow pos(x, y, d)$;
18 **end**
19 If $\psi$ has not been found then perform exhaustiveSearch($P$);
20 **end**

**Fig. 14** *Algorithm 6: VNS3 algorithm*

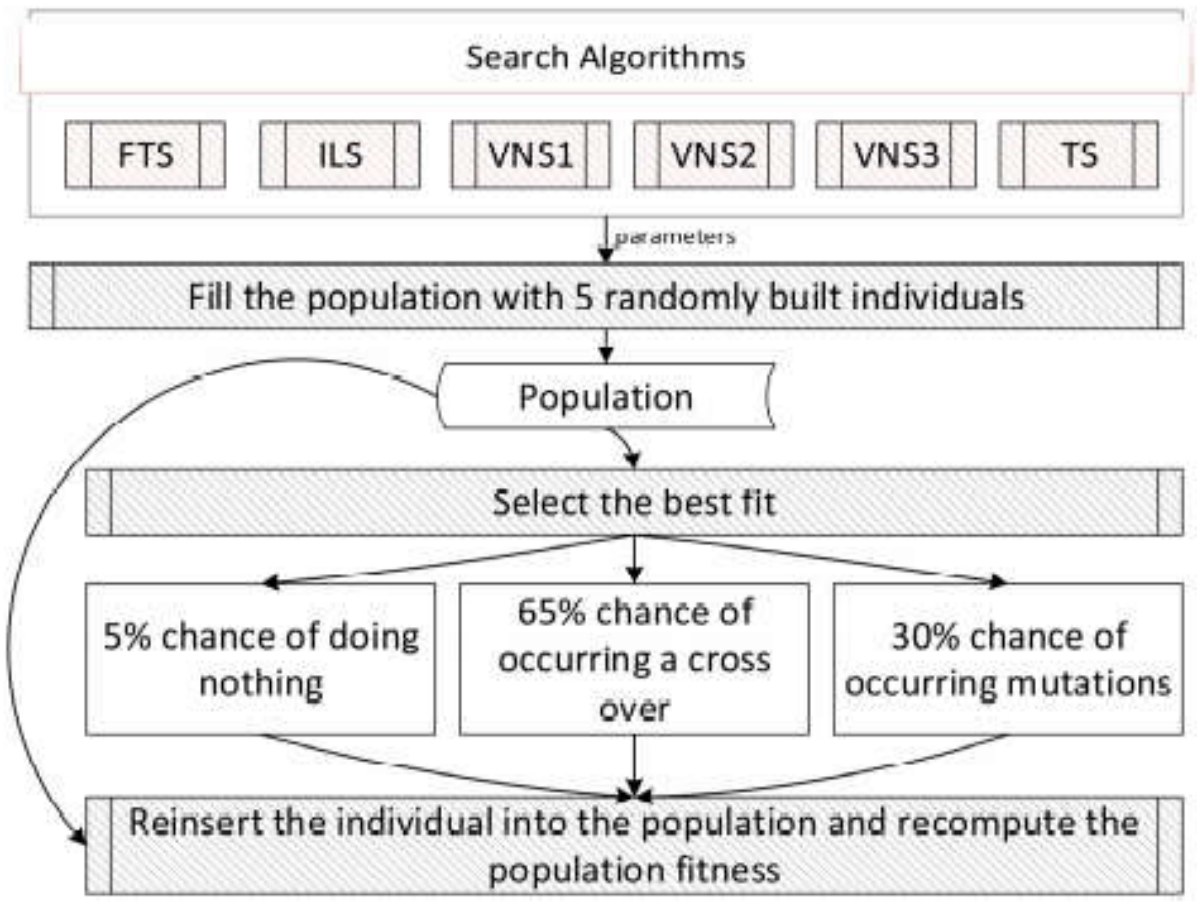

**Fig. 15** *Evolutionary algorithm used for parameter selection*

the lowest value. Cells highlighted in bold indicate occasions when FTS outperformed the remaining algorithms. Indeed, the FTS algorithm obtained better solutions in most cases (seven out of nine).

It is important to highlight that the average of $2 \times 10^7$ runs is very precise. The results in the exhaustive search column are close to the mean number of elements (i.e. the expected value using the law of large numbers) of the image or matrix, just as in the dice case $(1+2+3+4+5+6)/6 = 3.5$. That is, $1024 \times 1024$ elements have a total of 1,048,576 elements, and the expected value is shown as

$$\frac{\sum_{z=1}^{1024 \times 1024} z}{1024 \times 1024} = 524288.5 \quad (4)$$

The results presented in the exhaustive search column are very close to the expected value (524,288.5), and according to the law of large numbers, this indicates that the number of runs more than suffices for a fair comparison. In other words, this proves that FTS is in fact better than the remaining metaheuristics in the average case.

The best results (from rows 1 to 9 in Table 1) for each search algorithm are shown in Fig. 16. These images (from *a* to *f*) respect the same scale. The *y*-axis represents the number of visited elements, while the *x*-axis represents time or the number of iterations. The red lines represent the mean number of visited elements, as shown in rows 1–9 in Table 1.

In Fig. 16, FTS presents a more accentuated random pattern. This is due to one single fact: the EA of FTS algorithm did not converge in this case, while the EA selection process for the remaining search techniques converged. As previously stated, when the EA converges, the 'bad results' generated while the parameters of the algorithm were still adapting are not considered. In other words, the selected set of input parameters for the FTS algorithm was not optimal, and therefore, FTS can achieve even better results. The lower the red line is the better, as the algorithm visits a fewer number of elements on average. Despite the lack of EA convergence, which leads to the 'bad results' being considered, FTS has the lowest red line.

Instances containing $2048 \times 2048$ elements were also used in the experiments. The purpose of this investigation was to consider the gain offered by the FTS algorithm as the number of elements increases. In this case, seven restarts, instead of nine, of the EA were performed, to reduce the processing time burden. Each line of Table 2 aggregates the mean number of visited pixels over $6 \times 10^6$ iterations.

Processing $2048 \times 2048$ instances is significantly slower than processing $1024 \times 1024$ instances. Therefore, we reduced the number of runs from $2 \times 10^7$ to $6 \times 10^6$ times in this experiment. In both occasions, the algorithms ran non-stop for approximately an entire week in an Athlon X4 processor clocked at 2.8 GHZ.

In this case, the results were even better. FTS had lower performance than VNS3 once, against 2 loses in the $1024 \times 1024$ case. Moreover, FTS produced the best mean and also the lowest visits number (best) results. From this initial investigation, it would appear that the gain of FTS is more profound when the size of the input image increases.

### *4.1 Experiments with computed tomography*

In this subsection, a practical experiment using computed tomography (CT) images is performed. In a previous work [19], we proposed a method that automatically segments two types of cardiac fats: epicardial and mediastinal.

In summary, the human heart is enclosed by a concentric sac named epicardium. The fat inside this sac is called epicardial, while the fat attached to the outer surface of the heart is named mediastinal. The proposed method consisted of three steps: (i) image registration, (ii) feature extraction and pixel-oriented classification and (iii) image dilation, respectively. In the first step, which concerns the process of registration, the retrosternal area, highlighted by the red circle in Fig. 17, is used as the alignment

**Table 1** Mean amount of visited elements with 1024 × 1024 instances for nine different highly optimised sets of parameters

| Parameters | FTS | ILS | VNS1 | VNS2 | VNS3 | Tabu | Exhaustive |
|---|---|---|---|---|---|---|---|
| 1 | 523,816 | 727,471 | 516,026 | 512,383 | 511,741 | 512,608 | 524,418 |
| 2 | **419,963** | 524,220 | 516,491 | 498,042 | 522,845 | 497,559 | 524,481 |
| 3 | **423,162** | 524,741 | 523,402 | 519,265 | 513,192 | 500,301 | 523,730 |
| 4 | **389,414** | 524,114 | 521,230 | 519,995 | 456,498 | 519,974 | 523,829 |
| 5 | **446,812** | 524,517 | 523,149 | 522,597 | 473,727 | 514,937 | 524,052 |
| 6 | **486,439** | 536,265 | 522,339 | 514,460 | 522,591 | 514,794 | 524,008 |
| 7 | 429,398 | 523,948 | 522,265 | 517,942 | 382,250 | 503,134 | 524,656 |
| 8 | **439,929** | 523,738 | 515,134 | 522,474 | 519,494 | 522,363 | 524,515 |
| 9 | **475,382** | 523,728 | 524,175 | 511,724 | 501,179 | 511,281 | 524,780 |
| mean | **448,257** | 548,082 | 520,467 | 515,431 | 489,279 | 510,772 | 524,274 |
| lowest | 389,414 | 523,728 | 515,134 | 498,042 | **382,250** | 497,559 | 523,730 |

Bold cases indicate that FTS outperformed other metaheuristics. In the mean and lowest lines, bold indicates the best performance.

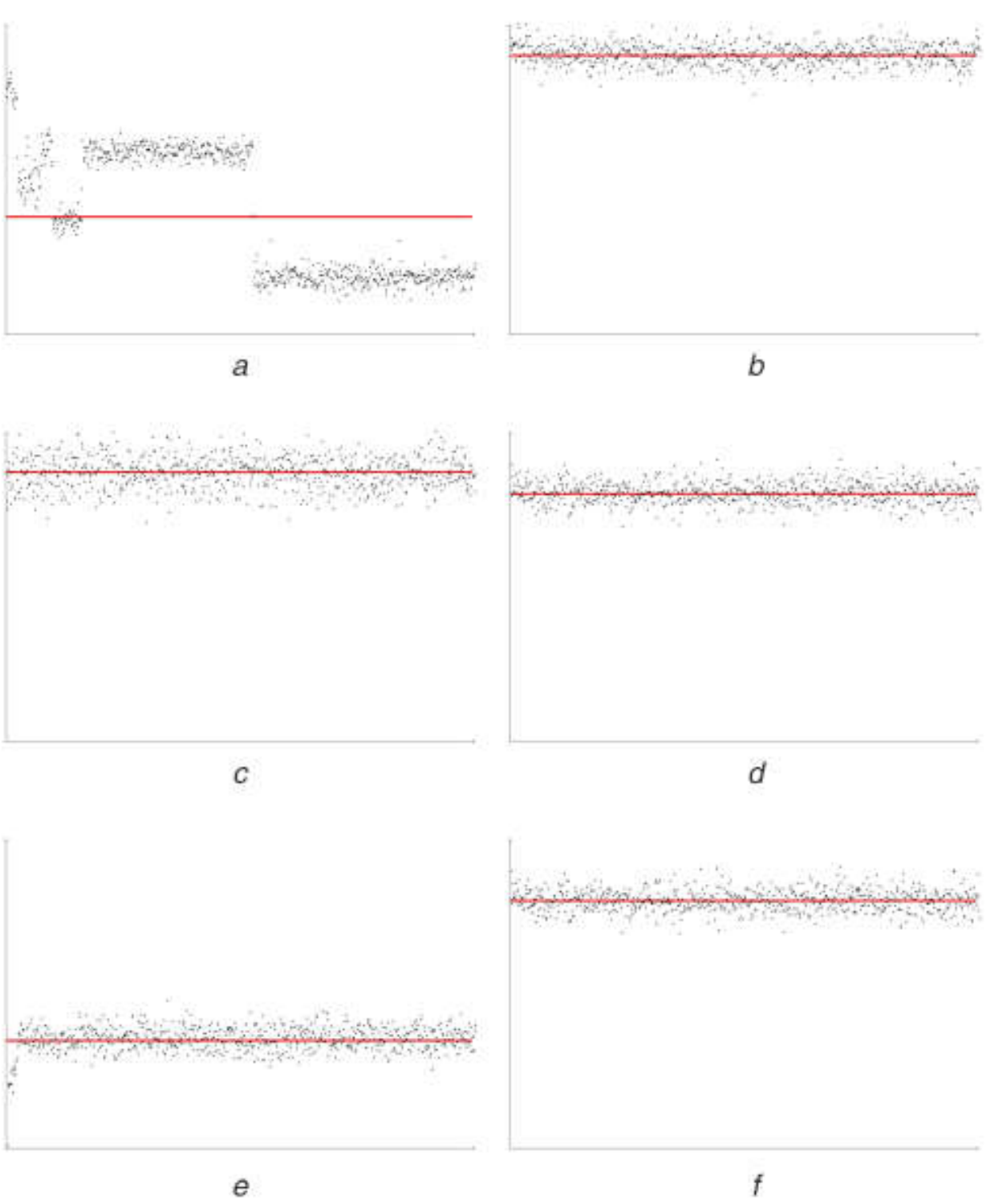

**Fig. 16** *Number of visited elements (y-axis) versus time (x-axis) for each algorithm. The red line illustrates the average number of visited elements (the lower the better)*

***(a)*** FTS, ***(b)*** ILS, ***(c)*** VNS1, ***(d)*** VNS2, ***(e)*** VNS3, ***(f)*** Tabu

**Table 2** Mean amount of visited elements with 2048 × 2048 instances for seven different highly optimised sets of parameters

| Parameters | FTS | ILS | VNS1 | VNS2 | VNS3 | Tabu | Exhaustive |
|---|---|---|---|---|---|---|---|
| 1 | **1,649,221** | 2,095,278 | 2,092,702 | 2,068,480 | 1,962,580 | 2,092,643 | 2,097,732 |
| 2 | **1,427,009** | 2,099,331 | 2,076,411 | 2,086,406 | 2,056,534 | 2,079,655 | 2,095,888 |
| 3 | 1,826,972 | 2,098,356 | 2,080,438 | 2,087,504 | 1,676,642 | 2,089,803 | 2,094,029 |
| 4 | **1,515,956** | 2,089,542 | 2,084,568 | 2,069,884 | 2,019,944 | 2,068,443 | 2,098,415 |
| 5 | **1,543,164** | 2,098,788 | 2,083,935 | 2,093,396 | 1,655,049 | 2,072,144 | 2,097,390 |
| 6 | **1,238,803** | 2,098,863 | 2,095,385 | 2,065,760 | 2,087,229 | 2,065,366 | 2,095,142 |
| 7 | **2,096,031** | 2,098,923 | 2,092,683 | 2,069,839 | 3,991,479 | 2,091,578 | 2,096,031 |
| mean | **1,613,879** | 2,097,011 | 2,086,588 | 2,077,324 | 2,207,065 | 2,079,947 | 2,096,375 |
| lowest | **1,238,803** | 2,089,542 | 2,076,411 | 2,065,760 | 1,655,049 | 2,065,366 | 2,094,029 |

Bold cases indicate that FTS outperformed other metaheuristics. In the mean and lowest lines, bold indicates the best performance.

reference/keypoint among different patients. The objective is to align different patients to a standard position. As different patients were registered using this methodology, this was defined as intersubject registration.

Since the retrosternal area is used as a common reference point among all patients, this region is automatically located in the image. Once the location of the retrosternal area is determined, this information is used to displace the image of the patient to a predefined standard location, therefore enabling alignment of structures across different patients and scanner manufacturers (Phillips, Siemens etc.).

The current state of the approach considers a brute force search, i.e. the exhaustive search approach presented in this work. That is, the retrosternal area is searched for in the entire image, and every possible location is evaluated. Essentially, it is assumed that the retrosternal area could be located anywhere in the image due to different acquisition practices and manufacturer protocols/ standards. A template that consists of the average of ten random retrosternal areas assists with the search. The template is displaced over the CT images, while the mean absolute error between the template and the corresponding section of the CT image is computed for each possible position of the template over the processed image. The position where this error score is the least is chosen as the correct location of the restrosternal area.

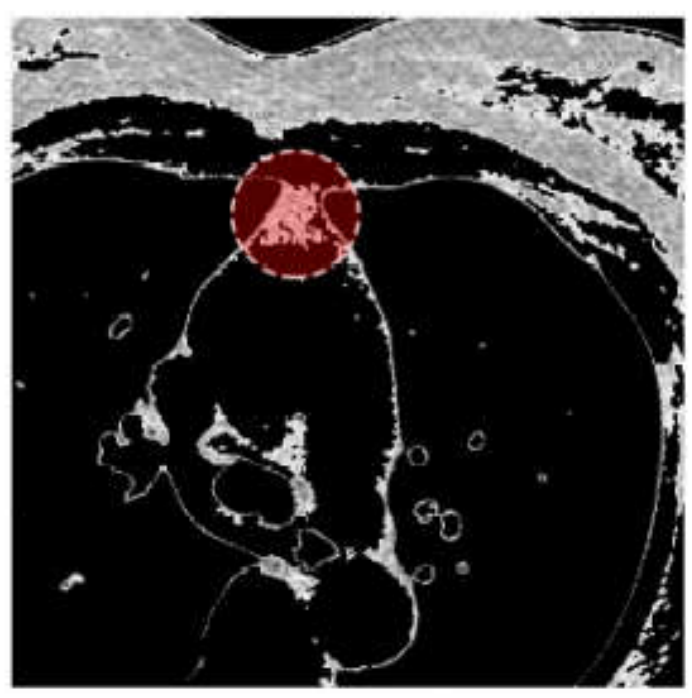

**Fig. 17** *Retrosternal area of a cardiac computed tomography image in the axial plane*

In this work, the previously used brute force registration approach is slightly modified [19]. Surrounding templates $\mu_1$ to $\mu_4$ are proposed to act as evidence elements, as depicted in Fig. 18, which shows a cardiac CT image of another randomly selected patient. In addition to computing the mean absolute error of the retrosternal template, the set of error metrics for each of these evidence templates is also computed. Therefore, while searching for the retrosternal area, one of these evidence areas may be found first. When this occurs, it is known, by definition, that the retrosternal area is located nearby and therefore this evidence information can be used to speed up the process of locating it.

In summary, we use template elements to bias the search in order to locate the main content faster (the retrosternal area – shown in Fig. 11). If the retrosternal area is detected first, the search stops. Otherwise, if one of the evidence elements is found, then the retrosternal area will be present nearby, subsequently triggering a local search, which is addressed differently by each of the presented algorithms. If no evidence nor main content is found, the search continues in a stochastic fashion. Both the evidence and the retrosternal areas are located using the presented templates, where an error score is computed at the evaluated position. In this approach, predefined thresholds indicate that the evidence region or the retrosternal area is found. That is, if the computed error score is below the threshold $\tau$, then the evaluated region is treated as an evidence region or as the retrosternal region.

Table 3 shows the relation of each algorithm (how faster) in comparison to the exhaustive search algorithm at the task of automatically locating the retrosternal area. Each table cell was calculated using a mean of 200 experiments, where different CT images were used for each case. A total of five runs were performed, which are shown in five different rows of the table (as the algorithms are stochastic, they may exhibit slightly different behaviour and run times).

It is possible to observe that FTS was 17.39% faster on average when compared to the exhaustive search approach. As a remark,

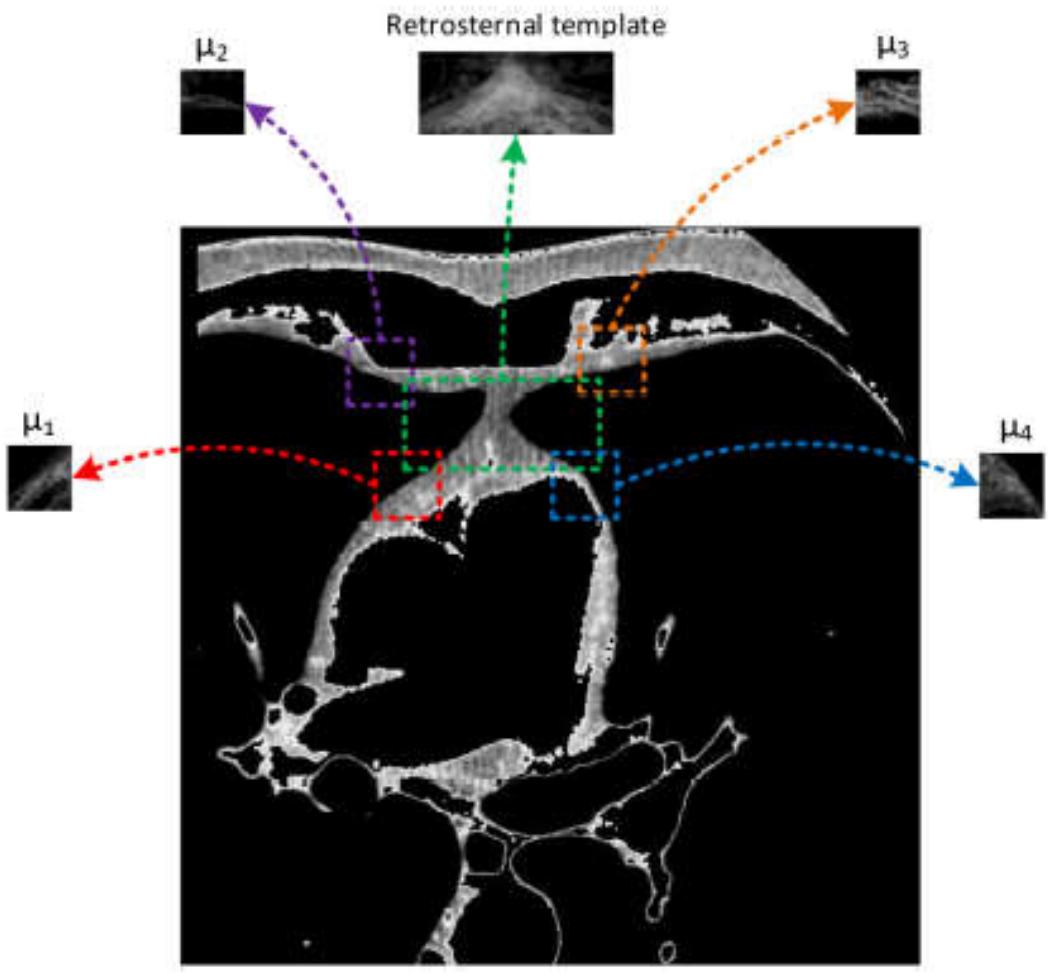

**Fig. 18** *Evidence elements surrounding the retrosternal area (search area). The evidence templates are the average of ten randomly selected images*

the exhaustive approach converged in 255 seconds on average. The second best search was VNS3, being 8.14% faster than the exhaustive search algorithm on average. FTS is expected to deliver higher margins of performance improvement if larger images/ search spaces are used. In this experiment, the CT images contain $512 \times 512$ pixels.

## 5 Conclusion

This work proposes a metaheuristic to be applied in content search in 2D discrete spaces or images. The results obtained with the proposed FTS metaheuristic outperformed all algorithms in comparison (ILS, VNS1-3, tabu and exhaustive). Dynamic instances of sizes $1024 \times 1024$ and $2048 \times 2048$ were generated and evaluated 'towards infinity'. An evolutionary algorithm was used to adjust the parameters of each search algorithm. The parameters of the evolutionary algorithm were carefully designed to deliver the expected outcomes. As a remark, EA took a long time to converge, and up to 2 days in certain occasions. The evolutionary algorithm was restarted nine times for the first experiment ($1024 \times 1024$) and seven times for the $2048 \times 2048$ case to avoid entrapment in local minima.

For each parameter selection using the evolutionary algorithm, each search algorithm was run ~20,000,000 times and the average of the results were compared (Tables 1 and 2). The FTS algorithm was better than the remaining in seven cases out of nine in the first experiment and in six times out of seven in the second experiment. In total, 4 months of continuous processing were dedicated to performing the experiments, which provide strong evidence that FTS is better than the state-of-the-art approaches in the average case.

This is the first work to provide practical solutions for the proposed SBE problem and to apply them in a real-world scenario in CT image registration. Future research will investigate the application of the FTS metaheuristic to several areas of visual computing, optimisation and data mining. Finally, all algorithms were implemented in Java and run over the JRE 7 in a single computer with an Athlon X4 processor clocked at 2.8 GHZ with 4 GB of RAM.

**Table 3** How many times faster each metaheuristic locates the retrosternal area in relation to the exhaustive approach

| Run | FTS | ILS | VNS1 | VNS2 | VNS3 | Tabu | Exhaustive |
|---|---|---|---|---|---|---|---|
| 1 | 16.32 | −4.10 | 1.16 | 2.67 | 9.51 | 3.28 | — |
| 2 | 14.51 | −6.06 | 0.92 | 1.39 | 8.52 | 3.55 | — |
| 3 | 21.02 | −5.79 | 0.94 | 2.27 | 8.56 | 1.67 | — |
| 4 | 19.25 | −5.56 | 0.83 | 1.55 | 7.01 | 4.61 | — |
| 5 | 15.89 | −4.73 | 0.75 | 2.62 | 7.10 | 3.06 | — |
| mean | 17.39 | −5.248 | 0.92 | 2.1 | 8.14 | 3.23 | — |

## 6 Acknowledgment

This work was supported by CAPES.